\documentclass[twocolumn,prl,superscriptaddress,showpacs,preprintnumbers,amsmath,amssymb]{revtex4}

\usepackage[ansinew,latin1]{inputenc}
\usepackage{psfrag}
\usepackage{array}
\usepackage{amssymb}
\usepackage{amsmath}
\usepackage{graphicx}% Include figure files

\begin{document}

\title{Comments on ``Limits on possible new nucleon monopole-dipole interactions from the spin relaxation rate of polarized $^3$He gas''}
\def\ILL{Institut Laue--Langevin, F--38042 Grenoble Cedex, France}
\def\Golub{North Carolina State University, Raleigh, NC, 27695-8202 USA}
\def\LPSC{LPSC, Universit\'e Joseph Fourier, CNRS/IN2P3, INPG, Grenoble, France}

\author{A.~K.~Petukhov}     	\affiliation{\ILL}
\author{G. Pignol}          	\affiliation{\LPSC}
\author{R. Golub}          	\affiliation{\Golub}
%\author{D. Jullien}          	\affiliation{\ILL}
%\author{K. H. Andersen}         \affiliation{\ILL}
%\author{E. Kats}          	\affiliation{\ILL}
%\author{R. Whitney}          	\affiliation{\ILL}

\date{\today}

\begin{abstract}
In the article ``Limits on possible new nucleon monopole-dipole interactions from the spin relaxation rate of polarized $^3$He gas'', 
new limits on short-range, Axion-like interactions are presented. 
In this comment it is shown that the theoretical treatement of the data overestimates the sensitivity of the proposed method. 
We provide the corrected limits. 
\end{abstract}

%\pacs{14.20.Dh, 11.30.Er, 11.30.Cp, 28.20.-v}
%pacs{}

\maketitle

The problem of relaxation in nuclear magnetic resonance due to field gradients has been discussed by many authors and continues to be a topic of current research. 
Recently attention has been focussed on this subject in connection with the search for new P,T violating forces mediated by light Axion-like particles \cite{Fu,Petukhov:2010dn,Pokotilovski}. 
It was recognized that new fundamental spin-dependent interaction between polarized atom of a gas and nucleons in the body of gas container may be treated as a usual interaction of the spin with magnetic/pseudo-magnetic field. 
The short-range character of such field results in a pseudo-magnetic field gradient near to the container walls which, in turn, leads to extra spin-relaxation of the polarized gas. 
Authors of \cite{Fu,Pokotilovski} explored this analogy in attempts to adapt well established theory \cite{Cates,McGregor,Golub} of  gas spin-relaxation in presence of uniform magnetic field gradient. 
In this way, a crucial question was raised: 
does the spin-relaxation rate induced by a short-range field follows from a simple averaging of the relaxation in uniform gradient \cite{Cates,McGregor} over the container volume? 
This question was answered negatively in \cite{Petukhov:2010dn} where a general solution of spin-relaxation in field with arbitrary spatial variation was obtained for one-dimentional geometry (expression (6)), 
the generalization for 3D geometry may be found in \cite{Clayton}. 
It has been shown \cite{Petukhov:2010dn}, expression (9), that in case of short-range pseudo-magnetic field $b({\bf r})$, the longitudinal relaxation $\Gamma_1$ does not follow from the volume averaging of the expression obtained for uniform gradient. 
A similar conclusion was obtained for the transverse relaxation time $\Gamma_2$. 
Let us consider a cubic cell of size $L$, with a $b$ field decaying exponentially from the top and bottom surfaces 
$b_z(z) = b_a \left( e^{-(L/2+z)/\lambda} - e^{-(L/2-z)/\lambda} \right)$. 
The transverse depolarization rate can be derived easily along the lines of \cite{Petukhov:2010dn}:
\begin{eqnarray}
\label{correct}
\Gamma_2 & = & (\gamma b_a)^2 \frac{\lambda^2}{D} \left( 1 + e^{-L/\lambda}\right)^2 \times \\ 
\nonumber
	& & \left(1+ \frac{1}{2} {\rm sech}^2(\frac{L}{2 \lambda}) - \frac{3 \lambda}{L} \tanh(\frac{L}{2 \lambda}) \right),
\end{eqnarray}
where $\gamma$ is the gyromagnetic ratio of the atoms and $D$ is the diffusion coefficient. 
Now, two limiting cases can be discussed. 
First, the case $\lambda \gg L$ represents the situation of an homogeneous field gradient of value $\partial_z b_z = b_a/\lambda$. 
In this case eq. (\ref{correct}) reduces to
\begin{equation}
\label{homogeneous}
\Gamma_2 \approx \frac{\gamma^2}{30} \left( \partial_z b \right)^2 \ \frac{L^4}{D}, 
\end{equation}
which coincides with the result \cite{McGregor} obtained for a spherical cell sitting in a magnetic field with a uniform gradient (up to a numerical factor of order unity due to different geometries). 
Moreover, in the previous work \cite{Petukhov:2010dn} exact agreement with \cite{McGregor} (eq. (26)) was found when a uniform gradient is substituded in the result for arbitrary field (eq. (6) in \cite{Petukhov:2010dn}). 
In the opposite case $\lambda \ll L$, which is the case of interest here, the formula (\ref{correct}) reduces to
\begin{equation}
\label{correctSimple}
\Gamma_2 \approx (\gamma b_a)^2  \frac{\lambda^2}{D}.
\end{equation}
We conclude that, when dealing with short range gradients, both longitudinal and transverse relaxation rates cannot be evaluated 
by taking the volume average of the square of the gradient and inserting the result in formula (61) in \cite{Cates} or (47) in \cite{McGregor}
as it is done in \cite{Fu,Pokotilovski}. 
Our analysis \cite{Petukhov:2010dn} shows the important influence of the spatial frequency spectrum of the perturbating field. 
This limits the validity of the previous theory \cite{Cates,McGregor} to the sole case of uniform gradients. 
%by the volume average of the corresponding expressions obtained for uniform gradient as it is done in \cite{Fu,Pokotilovski}.
This conclusion was further confirmed by MonteCarlo simulations which showed that the result (\ref{correctSimple}) is also valid in other geometries, up to numerical factors of order one 
(surprisingly, the result is valid even when the range of the interaction $\lambda$ is smaller than the atomic mean free path).

\begin{figure}
\includegraphics[width=0.87\linewidth]{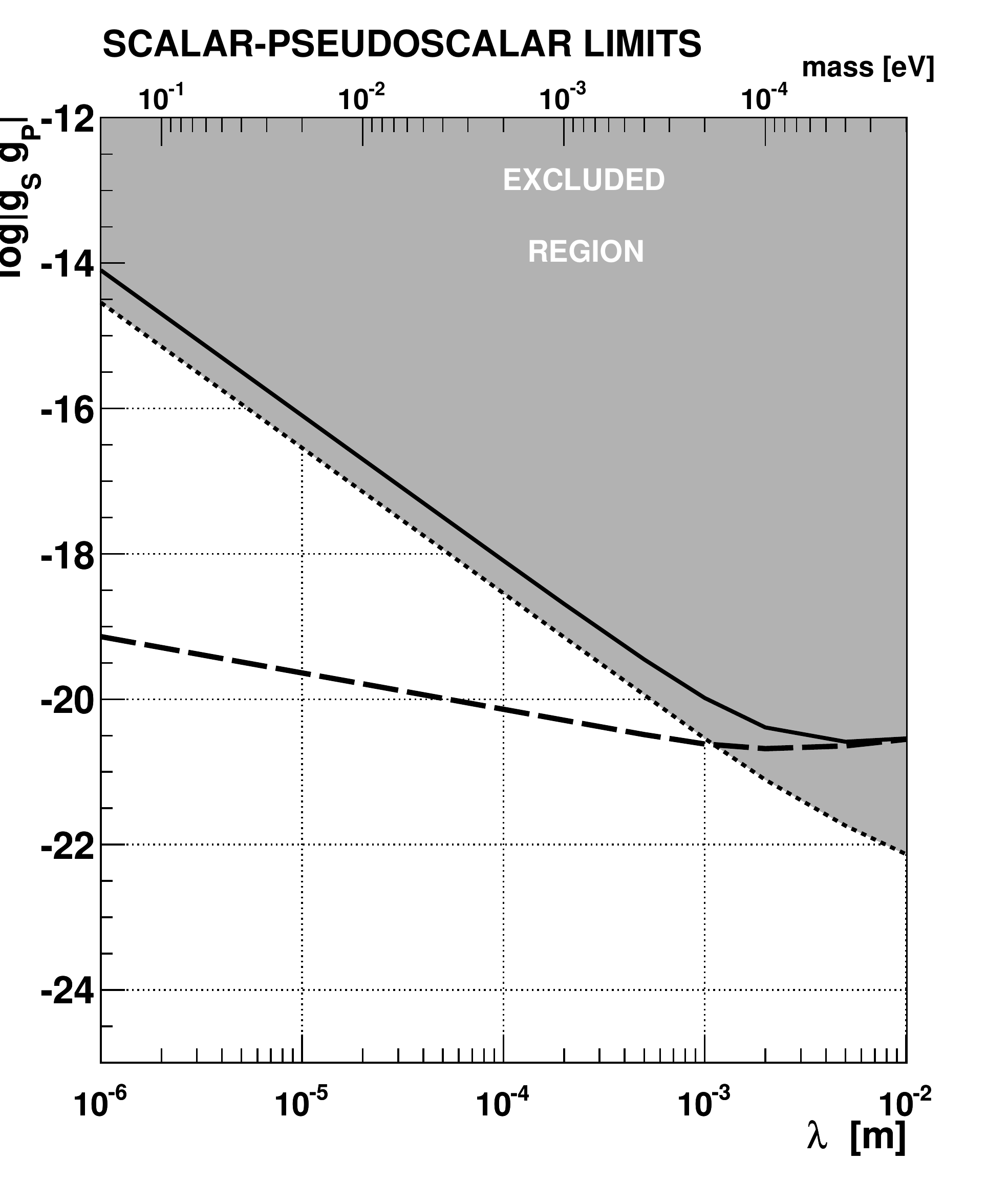}
\caption{
Constraints to the coupling constant product of axionlike particles to nucleons $g_s g_p$ as a function of the range $\lambda$ of the macroscopic interaction. 
Dashed line, from \cite{Fu} using transverse relaxation of high pressure $^3$He (incorrect), bold line, from \cite{Fu} (corrected). 
Dotted line, from \cite{Petukhov:2010dn} using transverse relaxation of low pressure $^3$He. 
} \label{Exclusion}
\end{figure}

Finally we present in fig. \ref{Exclusion} the corrected limits on new short-range, Axion-like interactions. 

We are greatful to E. Kats and R Whitney for valuable discussion. 

%-------------------------%

\end{document}